%% file: samplepaper.tex
\documentclass[graybox]{svmult}

\usepackage{type1cm}        
\usepackage{makeidx}         
\usepackage{graphicx}        
\usepackage{multicol}        
\usepackage[bottom]{footmisc}

\usepackage{newtxtext}       %
\usepackage{newtxmath}       

\usepackage{multirow}

\newcommand{\eg}{e.\,g., }
\newcommand{\ie}{i.\,e., }

\makeindex             

\begin{document}

\title*{Can Author Collaboration Reveal Impact? The Case of $h$-index}
\author{Giannis Nikolentzos \and George Panagopoulos \and Iakovos Evdaimon \and Michalis Vazirgiannis}
\authorrunning{Giannis Nikolentzos el al.}
\institute{Giannis Nikolentzos \at Athens University of Economics and Business, Athens, Greece, \email{nikolentzos@aueb.gr} \and George Panagopoulos \at \'Ecole Polytechnique, Palaiseau, France, \email{george.panagopoulos@polytechnique.edu} \and Iakovos Evdaimon \at Athens University of Economics and Business, Athens, Greece, \email{p3130059@aueb.gr} \and Michalis Vazirgiannis \at \'Ecole Polytechnique, Palaiseau, France, \email{mvazirg@lix.polytechnique.fr}}
%
%
\maketitle

\abstract*{Scientific impact has been the center of extended debate regarding its accuracy and reliability.
From hiring committees in academic institutions to governmental agencies that distribute funding, an author's scientific success as measured by $h$-index is a vital point to their career.
The objective of this work is to investigate whether the collaboration patterns of an author are good predictors of the author's future $h$-index.
Although not directly related to each other, a more intense collaboration can result into increased productivity which can potentially have an impact on the author's future $h$-index.
In this paper, we capitalize on recent advances in graph neural networks and we examine the possibility of predicting the $h$-index relying solely on the author's collaboration and the textual content of a subset of their papers.
We perform our experiments on a large-scale network consisting of more than $1$ million authors that have published papers in computer science venues and more than $37$ million edges.
The task is a six-months-ahead forecast, \ie what the $h$-index of each author will be after six months.
Our experiments indicate that there is indeed some relationship between the future $h$-index of an author and their structural role in the co-authorship network.
Furthermore, we found that the proposed method outperforms standard machine learning techniques based on simple graph metrics along with node representations learnt from the textual content of the author's papers.
}

\abstract{Scientific impact has been the center of extended debate regarding its accuracy and reliability.
From hiring committees in academic institutions to governmental agencies that distribute funding, an author's scientific success as measured by $h$-index is a vital point to their career.
The objective of this work is to investigate whether the collaboration patterns of an author are good predictors of the author's future $h$-index.
Although not directly related to each other, a more intense collaboration can result into increased productivity which can potentially have an impact on the author's future $h$-index.
In this paper, we capitalize on recent advances in graph neural networks and we examine the possibility of predicting the $h$-index relying solely on the author's collaboration and the textual content of a subset of their papers.
We perform our experiments on a large-scale network consisting of more than $1$ million authors that have published papers in computer science venues and more than $37$ million edges.
The task is a six-months-ahead forecast, \ie what the $h$-index of each author will be after six months.
Our experiments indicate that there is indeed some relationship between the future $h$-index of an author and their structural role in the co-authorship network.
Furthermore, we found that the proposed method outperforms standard machine learning techniques based on simple graph metrics along with node representations learnt from the textual content of the author's papers.
}

\section{Introduction}
Citation counts is undoubtedly the most widely used indicator of a paper's impact and success.
It is commonly believed that the more citations a paper has received, the higher its impact.
When it comes to authors, measuring impact becomes more complicated since an author may have published multiple papers in different journals or conferences.
Still the publication record of an author is in many cases the most important criterion for hiring and promotion decisions, and for awarding grants.
Therefore, institutes and administrators are often in need of quantitative metrics that provide an objective summary of the impact of an author's
publications.
Such indicators have been widely studied in the past years \cite{egghe2006theory,bornmann2008there,zhang2009index,wildgaard2014review}.
However, it turns out that not all aspects of an author's scientific contributions can be naturally captured by such single‐dimensional measures.
The most commonly used indicator is perhaps the $h$-index, a measure that was proposed by Jorge Hirsch \cite{hirsch2005index}.
The $h$-index measures both the productivity (\ie the number of publications) and the impact of the work of a researcher.
Formally, the $h$-index is defined as the maximum value of $h$ such that the given author has published $h$ papers that have each been cited at least $h$ times.
Since its inception in 2005, the $h$-index has attracted significant attention in the field of bibliometrics.
It is not thus surprising that all major bibliographic databases such as Scopus, the Web of Science and Google Scholar compute and report the $h$-index of authors.
One of the main appealing properties of the $h$-index is that it is easy to compute, however, the indicator also suffers from several limitations which have been identified in the previous years \cite{bornmann2005does,waltman2012inconsistency,roediger2006h}.

It is clear from the definition of $h$-index that it mainly depends on the publication record of the author.
On the other hand, there is no theoretical link between the $h$-index of an author and the collaborations the author has formed.
For instance, it is not necessary that the $h$-index of an author that has collaborated with many other individuals is high.
However, the above example is quite reasonable, and empirically, there could be some relation between $h$-index and co-authorship patterns \cite{katz1997much}.
In fact, it has been reported that co-authorship leads to increased scientific productivity (\ie number of papers published) \cite{zuckerman1967nobel,adams2005scientific,lee2005impact,ductor2015does,hu2014collaboration}.
For instance, some studies have investigated the relationship between productivity and the structural role of authors in the co-authorship network, and have reported that authors who publish with many different co-authors bridge communication and tend to publish larger amounts of papers \cite{eaton1999structural,kretschmer2004author}.
Since research productivity is captured by the $h$-index, co-authorship networks could potentially provide some insight into the impact of authors.
Some previous works have studied if centrality measures of vertices are related to the impact of the corresponding authors.
Results obtained from statistical analysis or from applying simple machine learning models have shown that in some cases, the impact of an author is indeed related to their structural role in the co-authorship network \cite{yan2009applying,abbasi2011identifying,bordons2015relationship}.

The goal of this paper is to uncover such connections between the two concepts, \ie $h$-index and the co-authorship patterns of an author.
Specifically, we leverage machine learning techniques and we study whether the collaboration patterns of an author are good indicators of the author's future impact (\ie their future $h$-index).
Predicting the future $h$-index of an author could prove very useful for funding agencies and hiring committees, who need to evaluate grant proposals and hire new faculty members, since it can give them insights about the future success of authors.
We treat the task as a regression problem and since the collaboration patterns of an author can be represented as a graph, we capitalize on recent advances in graph neural networks (GNNs) \cite{wu2020comprehensive}.
These models have been applied with great success to problems that arise in chemistry \cite{gilmer2017neural}, in social networks \cite{fan2019graph}, in natural language processing \cite{nikolentzos2020message}, and in other domains.
GNNs use the idea of recursive neighborhood aggregation.
Given a set of initial vertex representations, each layer of the model updates the representation of each vertex based on the previous representation of the vertex and on messages received from its neighbors.
In this paper, we build a co-authorship network whose vertices correspond to authors that have published papers in computer science journals and conferences.
We then train a GNN model to predict the future $h$-index of the vertices (\ie the authors).
More precisely, we propose two GNN variants tailored to the needs of the $h$-index prediction task.
Our results demonstrate that the structural role of an individual in the co-authorship network is indeed empirically related to their future $h$-index.
Furthermore, we find that local features extracted from the neighborhood of an author are very useful for predicting the author's future $h$-index.
On the other hand, the textual features generated from the author's published papers do not provide much information.
Overall, the proposed architectures achieve low levels of error, and can make relatively accurate predictions about the scientific success of authors.

The rest of this paper is organized as follows.
Section~\ref{sec:related_work} provides an overview of the related work.
In section~\ref{sec:data}, we present our dataset, while section~\ref{sec:preliminaries} introduces some preliminary concepts and gives details about representation learning in natural language processing and about graph neural networks.
Section~\ref{sec:methodology} provides a detailed description of our methodology.
Section~\ref{sec:experiments} evaluates the proposed graph neural network in the task of $h$-index prediction.
Finally, section~\ref{sec:conclusion} summarizes the work and presents potential future work.

\section{Related Work}\label{sec:related_work}
In the past years, a substantial amount of research has focused on determining whether information extracted from the co-authorship networks could serve as a good predictor of the authors' scientific performance.
For instance, McCarty \textit{et al.} investigated in \cite{mccarty2013predicting} which features extracted from the co-authorship network enhance most the $h$-index of an author.
The authors found that the number of co-authors and some features associated with highly productive co-authors are most related to the increase in the $h$-index of an author.
In another study \cite{heiberger2016choosing}, Heiberger and Wieczorek examined whether the structural role of researchers in co-authorship networks is related to scientific success, measured by the probability of getting a paper published in a high-impact journal.
The authors found that the maintenance of a moderate number of persistent ties, \ie ties that last at least two consecutive years, can lead to scientific success.
Furthermore, they report that authors who connect otherwise unconnected parts of the network are very likely to publish papers in high-impact journals.
Parish \textit{et al.} employed in \cite{parish2018dynamics} the $R$ index, an indicator that captures the co-authorship patterns of individual researchers, and studied the association between $R$ and the $h$-index.
The authors found that more collaborative researchers (\ie lower values of $R$) tend to have higher values of $h$-index, and the effect is stronger in certain fields compared to others.

Centrality measures identify the most important vertices within a graph, and several previous works have studied the relationship between these measures and research impact.
Yan and Ding investigated in \cite{yan2009applying} if four centrality measures (\ie closeness centrality, betweenness centrality, degree centrality, and PageRank) for authors in the co-authorship network are related to the author's scientific performance.
They found that all four centrality measures are significantly correlated with citation counts.
Abbasi \textit{et al.} also studied the same problem in \cite{abbasi2011identifying}.
Results obtained from a Poisson regression model suggest that the $g$-index (an alternative of the $h$-index) of authors is positively correlated with four of the considered centrality measures (\ie normalized degree centrality, normalized eigenvector centrality, average ties strength and efficiency).
Sarig{\"o}l \textit{et al.} examined in \cite{sarigol2014predicting} if the success of a paper depends on the centralities of its authors in the co-authorship network.
The authors showed that if a paper is represented as a vector where most of the features correspond to centralities of its authors, then a Random Forest classifier that takes these vectors as input can accurately predict the future citation success of the paper.
Bordons \textit{et al.} explored in \cite{bordons2015relationship} the relationship between the scientific performance of authors and their structural role in co-authorship networks. 
The authors utilized a Poisson regression model to explore how authors' $g$-index is related to different features extracted from the co-authorship networks.
The authors find that the degree centrality and the strength of links show a positive relationship with the $g$-index in all three considered fields.
Centrality measures cannot only predict scientific success, but can also capture the publication history of researchers. 
More specifically, Servia-Rodr{\'\i}guez \textit{\textit{et al.}} studied in \cite{servia2015evolution} if the research performance of an author is related to the number of their collaborations, and they found that if two authors have similar centralities in the co-authorship network, it is likely that they also share similar publication patterns.

Some other related studies have investigated the relationship between the number of authors and the success of a paper.
For instance, Figg \textit{et al.} analyzed in \cite{figg2006scientific} the relationship between collaboration and the citation rate of a paper.
The authors found that there exists a positive correlation between the number of authors and the number of citations received by a paper, for papers published in six leading journals.
Hsu and Huang studied in \cite{hsu2011correlation} whether collaboration leads to higher scientific impact.
They computed the correlation between the number of citations and the number of co-authors for papers published in eight different journals, and drew the same conclusions as Figg \textit{et al.} above.
Within each journal, there exists a positive correlation between citations and the number of co-authors, while single-authored articles receive the smallest number of citations.
In a different study \cite{abramo2011relationship}, Abramo \textit{et al.} examined how the international collaborations of Italian university researchers is associated with their productivity and scientific performance.
The authors found that the volume of international collaboration is positively correlated with productivity, while the correlation between the volume of international collaboration and the average quality of performed research is not very strong.

Our work is also related to mining and learning tasks in academic data such as collaboration prediction \cite{guns2014recommending,yan2014predicting} and prediction of scientific impact and success \cite{acuna2012predicting,weihs2017learning}.
When the scientific impact is measured by the author's $h$-index, the learning task boils down to predicting the author's future $h$-index \cite{dong2016can}, while a similar problem seeks to answer the question of whether a specific paper will increase the primary author's $h$-index \cite{dong2015will}.
In addition, some attempts have been made to predict the impact of an author based on its early indications and the use of graph mining techniques \cite{panagopoulos2017detecting}.

Finally, GNNs, as learning models, have also been applied to other types of graphs extracted from bibliographic data.
More specifically, some of the most common benchmark datasets to evaluate GNN methods are the following paper citation networks: Cora, PubMed and CiteSeer \cite{sen2008collective}.
These datasets are used to predict the field of the paper given the network and the paper's textual information (\ie abstract), in small scale.
In a larger scale, a recent graph transformer has been developed to perform link prediction on a heterogeneous network that includes papers, textual information extracted from papers, venues, authors, citations and fields \cite{hu2020heterogeneous}.
The edges represent different types of relations and thus an attention mechanism is utilized to separate between them and aggregate the representations of the nodes.
The final tasks evaluated are paper-field, paper-venue and author-paper matching which correspond to link prediction tasks.

\section{Data}\label{sec:data}
The Microsoft Academic Graph (MAG) is one of the largest publicly available bibliographic databases.
It contains more than 250 million authors and 219 million papers \cite{shen2018web}.
We decided to use MAG instead of other potential databases such as DBLP and AMiner for two reasons.
The first is that it has been found that MAG has a quite large coverage \cite{martin2020google}, and the $h$-index values of authors estimated from the data contained in MAG were closer to those provided by services like Google Scholar. 
The second is that MAG is more well-curated compared to other databases (\eg less conference names and scientific fields were missing).

\begin{figure}[t]
    \centering
    \includegraphics[width=.7\textwidth]{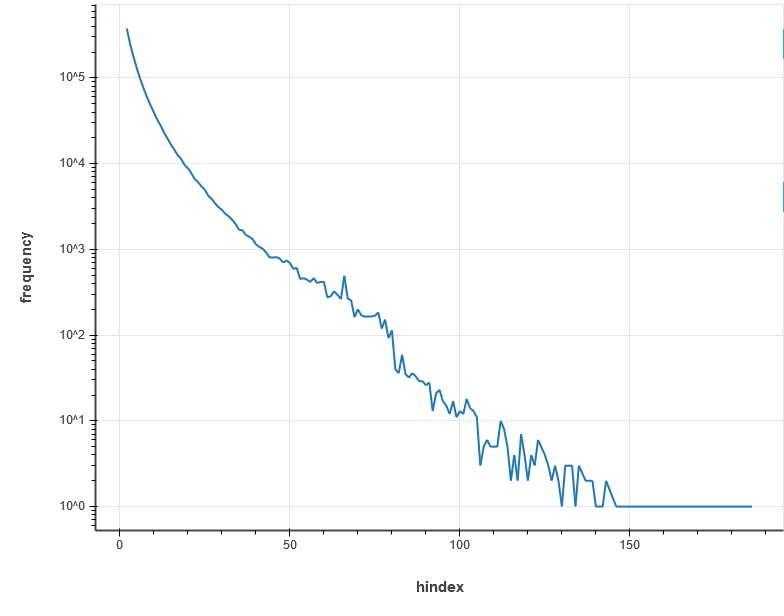}
    \caption{Distribution of $h$-index values of the authors contained in our dataset.}
    \label{fig:hindex}
\end{figure}

It should be noted that MAG required some pre-processing before predictive models could be applied, and due to its very large scale, this task turned out to be particularly challenging.
The initial pre-processing is delineated in \cite{panagopoulos2019scientometrics} and it included among others removing duplicate authors that are associated with a single, removing stopwords from the text using the NLTK library \cite{loper2002nltk}, and determining the scientific field of each author based on the ``field of study'' of their papers\footnote{\url{https://docs.microsoft.com/en-us/academic-services/graph/reference-data-schema}}.
The ``field of study'' of each paper is determined based on a machine learning pipeline that uses features such as the venues and the papers' textual content as described in \cite{sinha2015overview}. 
This allowed us to extract subgraphs of the network induced by authors that belong to a specific field of study. 
In this work, we focused on authors that have published papers which belong to the computer science field.
This graph was extracted and already used in another study on influence maximization \cite{panagopoulos2020influence}.
It should be noted that the emerging dataset was missing textual information of some papers.
If none of an author's papers contained text, we removed the corresponding node from the graph as we would not be able to leverage the heterogeneous information effectively.
The final graph consists of 1,503,364 nodes and 37,010,860 edges.
The weight of an edge is set equal to the number of papers two scholars have co-authored together.
To investigate if the proposed model can accurately predict the future $h$-index of authors, we extracted the $h$-index values of the authors from a more recent snapshot of MAG compared to the one we used to construct the co-authorship network (June 2019 vs. December 2018).
The $h$-index distribution of the authors is illustrated in Figure~\ref{fig:hindex}.
Note that the values on the vertical axis are in logarithmic scale.
As we see, it verifies the well-known power law distribution inherent in many real-world networks \cite{faloutsos1999power}.

\section{Preliminaries}\label{sec:preliminaries}
In this section, we first define our notation, we then present the different graph metrics that we employed, and we finally introduce the concepts of word embeddings and graph neural networks.

\subsection{Notation}
Let $G = (V,E)$ be an undirected and unweighted graph consisting of a set $V$ of vertices and a set $E$ of edges between them.
We will denote by $n$ the number of vertices and by $m$ the number of edges.
The neighbourhood $\mathcal{N}(v)$ of vertex $v$ is the set of all vertices adjacent to $v$.
Hence, $\mathcal{N}(v) = \{u : (v,u) \in E\}$ where $(v, u)$ is an edge between vertices $v$ and $u$ of $V$.
We denote the degree of vertex $v$ by $d_v = |\mathcal{N}(v)|$.
A graph $G$ can be represented by its adjacency matrix $\mathbf{A}$.
The $(i,j)^{\text{th}}$ entry of $\mathbf{A}$ is $w_{ij}$ if the edge $(v_i, v_j)$ between vertices $v_i$ and $v_j$ exists and its weight is equal to $w_{ij}$, and $0$ otherwise.
An attributed graph is a graph whose vertices are annotated with continuous multidimensional attributes.
We use $\mathbf{X} \in \mathbb{R}^{n \times d}$ where $d$ is the number of attributes to denote the graph's vertex information matrix with each row representing the attribute of a vertex. 

\subsection{Graph Metrics}
One of the most prominent ways to estimate the scientific impact of an author is based on the author's position in the academic collaboration network \cite{panagopoulos2017detecting}.
The position can be estimated through multiple network science centralities and metrics developed to capture different dimensions of a node's impact on the graph.
In this work, we utilize a number of them in order to compare them with the proposed approach and evaluate their usefullness in our framework. 
\begin{itemize}
    \item \textbf{Degree}: The sum of the weights of edges adjacent to a vertex.
    Since the co-authorship network is undirected, we can not compute the in-degree and out-degree.
    \vspace{.1cm}
    \item \textbf{Degree centrality}: It is the normalized degree of a vertex, \ie the number of neighbors of the vertex divided by the maximum possible number of neighbors (\ie $n-1$).
    This measure does not take into account the weights of the edges.
    \vspace{.1cm}
    \item \textbf{Neighbor's average degree}: The average degree of the neighborhood of a vertex:
    \begin{equation}
        neig\_avg\_deg(v) = \frac{1}{|\mathcal{N}(v)|}\sum_{u \in \mathcal{N}(v)} deg(u)
    \end{equation}
    where $deg(u)$ is the degree of vertex $u$.
    This measure also ignores edge weights.
    \vspace{.1cm}
    \item \textbf{PageRank}:
    Pagerank is an algorithm that computes a ranking of the vertices in a graph based on the structure of the incoming egdes.
    The main idea behind the algorithm is that a vertex spreads its importance to all vertices it links to:
    \begin{equation}
        PR(v) = \sum_{u \in \mathcal{N}(v)} \frac{PR(u)}{deg(u)}
    \end{equation}
    In practice, the algorithm computes a weighted sum of two matrices, \ie the column stochastic adjacency matrix and an all-ones matrix.
    Then, the pagerank scores are contained in the eigenvector associated with the eigenvalue 1 of that matrix \cite{page1999pagerank}.
    \vspace{.1cm}
    \item \textbf{Core number}:
    A subgraph of a graph $G$ is defined to be a $k$-core of $G$ if it is a maximal subgraph of $G$ in which all vertices have degree at least $k$ \cite{malliaros2020core}.
    The core number $c(v)$ of a vertex $v$ is equal to the highest-order core that $v$ belongs to.
    \vspace{.1cm}
    \item \textbf{Onion layers}: 
    The onion decomposition is a variant of the $k$-core decomposition \cite{hebert2016multi}.
    \vspace{.1cm}
    \item \textbf{Diversity coefficient}:
    The diversity coefficient is a centrality measure based on the Shannon entropy.
    Given the probability of selecting a node's neighbor based on its edge weight $p_{u,v} = \frac{w_{v,u}}{\sum_{u \in N(v)} w_{v,u}}$, the diversity of a vertex is defined as the (scaled) Shannon entropy of the weights of its incident edges.
    \begin{equation}
        D(v) = \frac{-\sum_{u \in \mathcal{N}(v)} ( p_{v,u} \log(p_{v,u}))}{\log(|\mathcal{N}(v)|)}
    \end{equation}
    \vspace{.1cm}
    \item \textbf{Community-based centrality}:
    This centrality measure calculates the importance of a vertex by considering its edges towards the different communities \cite{tulu2018identifying}.
    Let $d_{v,c}$ be the number of edges between vertex $v$ and community $c$ and $n_c$ the size of community $c$ retrieved by modularity optimization.
    The metric is then defined as follows:
    \begin{equation}
        CB_v = \sum_{c \in C} d_{v,c} \frac{n_c}{n}
    \end{equation}   
    \vspace{.1cm}
    \item \textbf{Community-based mediator}:
    The mediator centrality takes into consideration the role of the vertex in connecting different communities \cite{tulu2018identifying}, where the communities are again computed by maximizing modularity. 
    The metric relies on the percentages of the node's weighted edges that lie in its community relative to its weighted degree and the corresponding percentage for edges on different communities.
    To compute it, we first calculate the internal density of vertex $v$ as follows:
    \begin{equation}
        p^{c_v}_v = \frac{
        \sum_{u \in \mathcal{N}(v) \cup c_v}{w_{v,u}}}{deg(v)}
    \end{equation}
    where $c_v$ is the community to which $v$ belongs and can be replaced with the other communities to obtain the respective external densities.
    Given all densities, we can calculate the entropy of a vertex as follows:
    \begin{equation}
    H_v =  -p^{c_v}_v \log(p^{c_v}_v) - \sum_{c' \in C \setminus c_v} p^{c'}_v \log(p^{c'}_v)
    \end{equation}
    where $C$ is the set that contains all the communities.
    Finally, we compute the community mediator centrality: 
    \begin{equation}
    CM_v = H_v \frac{deg(v)}{\sum_{u \in \mathcal{N}(v)} deg(u) }   
    \end{equation}
\end{itemize}

\subsection{Text Representation Learning}
In the past years, representation learning approaches have been applied heavily in the field of natural language processing.
The Skip-Gram model \cite{mikolov2013distributed} is undoubtedly one of the most well-established methods for generating distributed representations of words.
Skip-Gram is a neural network comprising of one hidden layer and an output layer, and can be trained on very large unlabeled datasets.
In our setting, let $P_v$ denote a set that contains the abstracts of some papers published by author $v$.
Let also $W$ denote the vocabulary of all the abstracts contained in $\bigcup\limits_{v \in V} P_v$
.
Then, to learn an embedding for each word $w \in W$, our model is trained to minimize the following objective function:
\begin{equation}
    \mathcal{L} = \sum_{d \in P_v} \, \sum_{w_i\in d} \, \sum_{\substack{w_j \in \{w_{i-c}, \ldots, w_{i+c} \} \\ w_j \neq w_i}}{\log \big( p(w_j|w_i) \big)}
    \label{eq:w2v_loss}
\end{equation}
\begin{equation}
p(w_j|w_i) = \frac{\exp (\mathbf{v}_{w_i}^\top \, \mathbf{v}_{w_j}')}{\sum_{w \in W} \exp (\mathbf{v}_{w_i}^\top \, \mathbf{v}_w')}
\label{eq:softmax}
\end{equation}
where $c$ is the training context, $\mathbf{v}_{w_i}^\top$ is the row of matrix $\mathbf{H}$ that corresponds to word $w_i$, and $\mathbf{v}_{w_j}'$ is the column of matrix $\mathbf{O}$ that corresponds to $w_j$.
Matrix $\mathbf{H} \in \mathbb{R}^{|W|\times d}$ is associated with the hidden layer, while matrix $\mathbf{O} \in \mathbb{R}^{d \times |W|}$ is associated with the output layer.
Both these matrices are randomly initialized and are trained using the loss function defined above. 
The embeddings of the words are contained in the rows of matrix $\mathbf{H}$.
The main intuition behind the representations learnt by this model is that if two words $w_i$ and $w_j$ have similar contexts (\ie they both co-occur with the same words), they obtain similar representations.
In real scenarios, the above formulation is impractical due to the large vocabulary size, and hence, a negative sampling scheme is usually employed \cite{goldberg2014word2vec}.
The training set is built by generating word-context pairs, \ie for each word $w_i$ and a training context $c$, we create pairs of the form $(w_i,w_{i-c}), \ldots, (w_i,w_{i-1}),(w_i,w_{i+1}),(w_i,w_{i+c})$.

Note that since the abstract of a paper contains multiple words, we need to aggregate the embeddings of the words to derive an embedding for the entire abstract.
To this end, we simply average the representations of the words.
Furthermore, an author may have published multiple papers, and to produce a vector representation for the author, we again compute the average of the representations of (the available subset of) their papers:
\begin{equation}
    \mathbf{h}_v = \frac{1}{|P_v|} \sum_{d \in P_v} \frac{1}{|d|} \sum_{w \in d} \mathbf{v}_w
    \label{eq:author_rep}
\end{equation}

\subsection{Graph Neural Networks}
In the past years, GNNs have attracted a lot of attention in the machine learning community and have been successfully applied to several problems.
Most of the proposed GNNs share the same basic idea, and can be reformulated into a single common framework \cite{gilmer2017neural}.
These models employ a message passing procedure, where vertices update their representations based on their previous representation and on messages received from their neighbors.
While the concept of message passing over graphs and that of GNNs have been around for many years \cite{zhu2002learning,sperduti1997supervised,scarselli2008graph}, GNNs have only recently become a topic of intense research mainly due to the recent advent of deep learning.

A GNN model consists of a series of neighborhood aggregation layers.
As already mentioned, each one of these layers uses the graph structure and the node feature vectors from the previous layer to generate new representations for the nodes.
The feature vectors are updated by aggregating local neighborhood information.
Let $\mathbf{h}_v^{(0)} \in \mathbb{R}^d$ denote the initial feature vector of vertex $v$, and suppose we have a GNN model that contains $T$ neighborhood aggregation layers.
In the $t$-th neighborhood aggregation layer ($t > 0$), the hidden state $\mathbf{h}_v^{(t)}$ of a node $v$ is updated as follows:
\begin{equation}
    \begin{split}
        \mathbf{m}_v^{(t)} &= \textsc{AGGREGATE}^{(t)}  \Bigl( \Bigl\{ \mathbf{h}_u^{(t-1)} \vert  u \in \mathcal{N}(v) \Bigr\} \Bigr) \\
        \mathbf{h}_v^{(t)} &= \textsc{COMBINE}^{(t)}  \Bigl( \mathbf{h}_v^{(t-1)}, \mathbf{m}_v^{(t)}  \Bigr)
    \end{split}
    \label{eq:gnn_general}
\end{equation}
By defining different $\textsc{AGGREGATE}^{(t)}$ and $\textsc{COMBINE}^{(t)}$ functions, we obtain a different GNN variant.
For the GNN to be end-to-end trainable, both functions need to be differentiable.
Furthermore, since there is no natural ordering of the neighbors of a node, the $\textsc{AGGREGATE}^{(t)}$ function must be permutation invariant.
The node feature vectors $\mathbf{h}_v^{(T)}$ of the final neighborhood aggregation layer are usually passed on to a fully-connected neural network to produce the output.

\section{Methodology}\label{sec:methodology}
In this section, we give details about the two graph neural networks that we employed and we present their exact architecture.

\subsection{Proposed Architecture}
\begin{figure}[t]
    \centering
    \includegraphics[width=.8\textwidth]{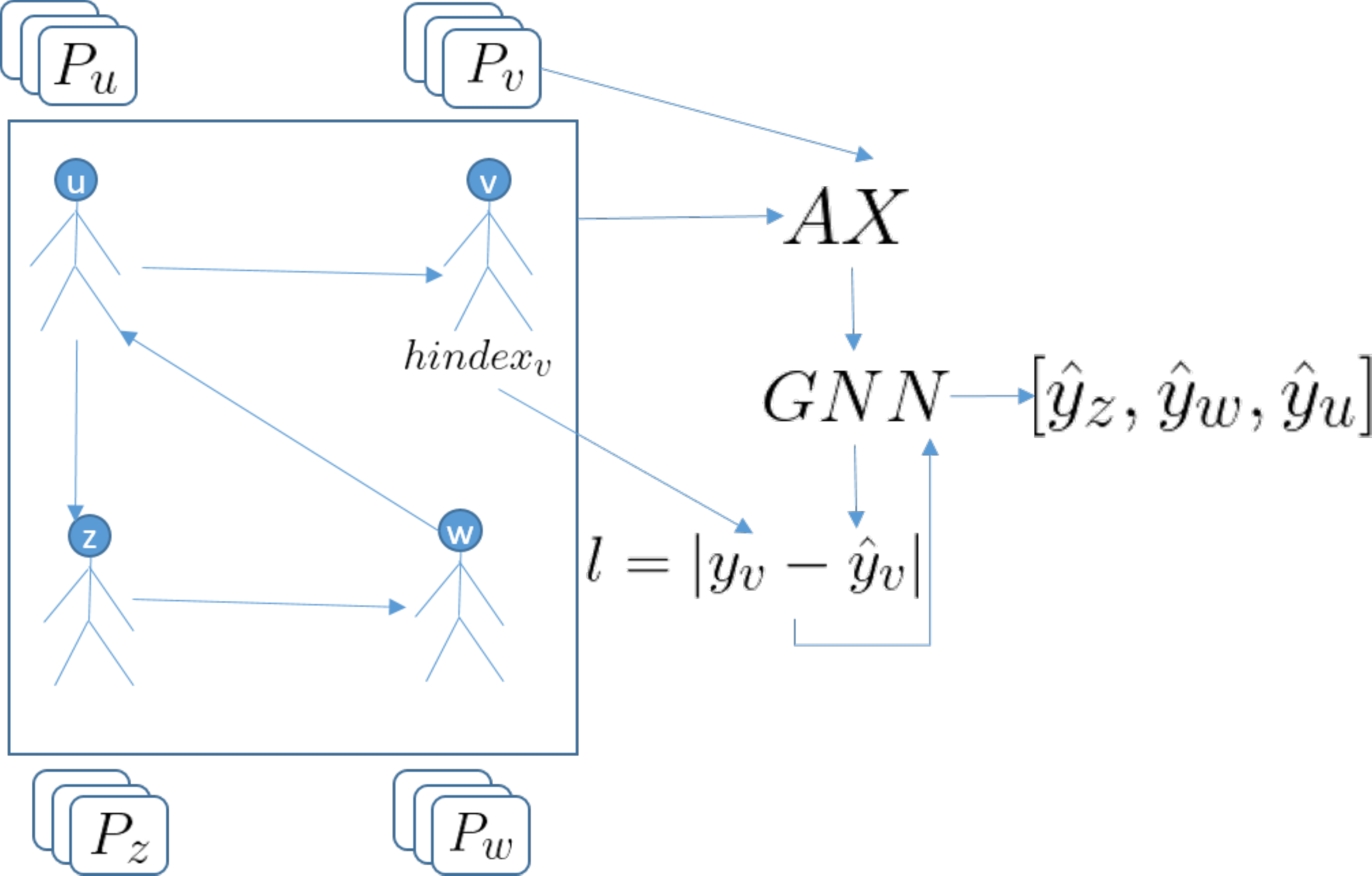}
    \caption{An overview of the proposed model for a graph of 4 nodes ($u,v,y,w$). The graph is represented by its adjacency matrix, while the textual and graph-related features of the nodes are stored in matrix $\mathbf{X}$. These two matrices are passed on to a GNN which predicts the authors' future $h$-index which are then compared against the actual $h$-index values to compute the loss.}
    \label{fig:scheme}
\end{figure}

We propose two different GNN variants which differ from each other only in terms of the employed message passing procedure.
Both models merge the $\textsc{AGGREGATE}$ and $\textsc{COMBINE}$ functions presented above into a single function.
A high-level illustration of the proposed approach is shown in Figure~\ref{fig:scheme}.

\paragraph{\textbf{GNN}.}
Given the aforementioned co-authorship graph $G = (V,E)$ where vertices are annotated with feature vectors $\mathbf{h}_v^{(0)} \in \mathbb{R}^d$ stemming from the learnt representations of the author's papers and/or the graph metrics, each neighborhood aggregation layer of the first model updates the representations of the vertices as follows:
\begin{equation}
    \mathbf{h}_v^{(t)} = \textsc{ReLU} \Bigl( \mathbf{W}^{(t)} \, \mathbf{h}_v^{(t-1)} + \sum_{u \in \mathcal{N}(v)} \mathbf{W}^{(t)} \, \mathbf{h}_u^{(t-1)} \Bigr)
    \label{eq:vector_1}
\end{equation}
where $\mathbf{W}^{(t)}$ is the matrix of trainable parameters of the $t^{\text{th}}$ message passing layer.
In matrix form, the above is equivalent to:
\begin{equation}
    \mathbf{H}^{(t)} = \textsc{ReLU} \Bigl( \tilde{\mathbf{A}} \, \mathbf{H}^{(t-1)} \, \mathbf{W}^{(t)} \Bigr)
    \label{eq:matrix_1}
\end{equation}
where $\tilde{\mathbf{A}} = \mathbf{A} + \mathbf{I}$.
Note that in both Equation~(\ref{eq:vector_1}) and~(\ref{eq:matrix_1}), we omit biases for clarity of presentation.
The above message passing procedure is in fact similar to the one of the GIN-0 model \cite{xu2019powerful}.

\paragraph{\textbf{GCN}.}
With regards to the second variant, its node-wise formulation is given by:
\begin{equation}
    \mathbf{h}_v^{(t)} = \textsc{ReLU} \Bigl( \mathbf{W}^{(t)} \, \frac{1}{1+d_v} \mathbf{h}_v^{(t-1)} + \sum_{u \in \mathcal{N}(v)} \mathbf{W}^{(t)} \, \frac{1}{\sqrt{(1+d_v)(1+d_u)}} \mathbf{h}_u^{(t-1)} \Bigr)
    \label{eq:vector_2}
\end{equation}
where $\mathbf{W}^{(t)}$ is the matrix of trainable parameters of the $t^{\text{th}}$ message passing layer.
In matrix form, the above is equivalent to:
\begin{equation}
    \mathbf{H}^{(t)} = \textsc{ReLU} \Bigl( \hat{\mathbf{A}} \, \mathbf{H}^{(t-1)} \, \mathbf{W}^{(t)} \Bigr)
    \label{eq:matrix_2}
\end{equation}
where $\hat{\mathbf{A}} = \tilde{\mathbf{D}}^{-\frac{1}{2}} \, \tilde{\mathbf{A}} \, \tilde{\mathbf{D}}^{-\frac{1}{2}}$ and $\tilde{\mathbf{D}}$ is a diagonal matrix such that $\tilde{\mathbf{D}}_{ii} = \sum_{j=1}^n \tilde{\mathbf{A}}_{ij}$.
Note that in both Equation~(\ref{eq:vector_2}) and~(\ref{eq:matrix_2}), we omit biases for clarity of presentation.
The above message passing procedure is in fact the one employed by the GCN \cite{kipf2017semi}.

It should be mentioned that matrix $\hat{\mathbf{A}}$ corresponds to the symmetrically normalized adjacency matrix with self-loops.
Normalization is mainly applied to avoid numerical instabilities and exploding/vanishing gradients associated with deep neural networks.  
One should note that by normalizing, we capture the distribution of the representations of the neighbors of a vertex, but not the exact multiset of representations.
Therefore, we could potentially lose the ability to distinguish between authors that are connected with many other researchers and authors that are connected only with a few other researchers \cite{xu2019powerful}. 
Note that besides the above two message passing schemes, we also tried using GAT-like attention \cite{velivckovic2018graph} in early experiments, without obtaining better results.

\paragraph{\textbf{Multi-hop information}.}
Inspired by Jumping Knowledge Networks \cite{xu2018representation}, instead of using only 
the final vertex representations $\mathbf{h}_v^{(T)}$ (\ie obtained after $T$ message passing steps), we also use the representations of the earlier steps $\mathbf{h}_v^{(1)},\ldots,\mathbf{h}_v^{(T-1)}$.
Note that as one iterates, vertex representations capture more and more global information.
However, retaining more local, intermediary information might be useful too.
Thus, we concatenate the representations produced at the different steps, finally obtaining $\mathbf{h}_v =[\mathbf{h}_v^{(1)}||\mathbf{h}_v^{(2)}||\ldots||\mathbf{h}_v^{(T)}]$.
These vertex representations are then passed on to one or more fully-connected layers to produce the output.

\section{Experimental Evaluation}\label{sec:experiments}
In this section, we first present the baselines against which we compared the proposed approach.
We next give details about the employed experimental protocol.
We last report on the performance of the different approaches and discuss the obtained results.

\subsection{Baselines}
We compare the proposed models against the following learning algorithms: (1) Lasso, a regression algorithm that performs both variable selection and regularization  \cite{tibshirani2011regression}, (2) SGDRegressor, a linear model which is trained by minimizing a regularized empirical loss with SGD, (3) XGBoost, a scalable end-to-end tree boosting system \cite{chen2016xgboost}, and a multi-layer perceptron (MLP).
All the above models expect each input sample to be in the form of a vector.
Given an author, the above vector is produced by the features extracted from the textual content of the author's papers and/or the features extracted from the co-authorship network.

\subsection{Experimental Setup}
In order to test the semi-supervised generalization capabilities of our model, we experimented with a $20/80$ training/test split. 
For all algorithms, we standardize the input features by removing the mean and scaling to unit variance.

With regards to the hyperparameters of the proposed models, we use $2$ message passing layers.
The hidden-dimension size of the message passing layers is set to $32$ and $64$, respectively.
In the case of the first model (\ie GNN), batch normalization is applied to the output of every message passing layer.
The representations produced by the second message passing layer are passed on to a multi-layer perceptron with one hidden layer of dimensionality $64$.
All dense layers use $\textsc{ReLU}$ activation. 
The dropout rate is set equal to $0.1$.
To train all models, we use the Adam optimizer with learning rate of $0.01$.
We set the number of epochs to $300$.
The best model is chosen based on a validation experiment on a single $90\%$ - $10\%$ split of the training data and is stored into disk.
At the end of training, the model is retrieved from the disk, and we use it to make predictions for the test instances.
The MLP contains a hidden layer of dimensionality $64$.
All its remaining hyperparameters (\eg dropout rate, activation function, etc.) take values identical to those of the two GNNs.
For lasso, the weight of regularization is set to $1$.
For SGDRegressor, we use an $l_2$ regularizer with weight $0.0001$.
The initial learning rate is set to $0.01$ and the number of epochs to $1000$.
For XGBoost, we use the default values of the hyperparameters.

\subsection{Results and Discussion}

\begin{table}[t]
    \centering
    \def\arraystretch{1.4}
    \begin{tabular}{l|c|c|c|c|c|c} \hline
    \multirow{2}{*}{Method} & \multicolumn{2}{|c|}{Text Features} & \multicolumn{2}{|c|}{Graph Features} & \multicolumn{2}{|c}{All Features} \\ \cline{2-7}
    & MAE & MSE & MAE & MSE & MAE & MSE \\ \hline
    Lasso & 4.99 & 66.91 & 3.28 & 29.51 & 3.28 & 29.51 \\
    SGDRegressor & 8.48 & 112.20 & 6.20 & 78.38 & 8.01 & 120.91 \\
    XGBoost & 4.22 & 64.43 & 3.04 & 34.83 & 2.91 & \textbf{21.04} \\
    MLP & 4.10 & 59.77 & \textbf{2.62} & \textbf{22.46} & 2.59 & 21.44 \\ \hline
    GCN & \textbf{4.05} & \textbf{59.45} & 2.68 & 24.32 & \textbf{2.57} & 21.29 \\
    GNN & 4.07 & 60.00 & 2.66 & 23.82 & 2.58 & 21.85 \\ \hline
    \end{tabular}
    \caption{Performance of the different methods in the $h$-index prediction task.}
    \label{tab:results}
\end{table}

The performance of the different models is illustrated in Table~\ref{tab:results}. 
We report the mean absolute error (MAE) and the mean squared error (MSE) of the different approaches.
There are three different sets of features passed on to each algorithm (features extracted from graph, features extracted from text or from both).
With regards to the performance of the different approaches, we first observe that neural network models outperform the other approaches in all settings and by considerable margins, except for one case.
We hypothesize that this stems from the inherent capability of neural networks to detect meaningful and useful patterns in large amounts of data.
In this case, though semi-supervised, the training set still consists of more than 300,000 samples, which allows for effective use of the representational power of neural network models.
On the other hand, we see that XGBoost performs better in terms of MSE in one setting since it optimizes the MSE criterion, while the neural architectures are trained by minimizing a MAE objective function, as also shown in Figure~\ref{fig:scheme}. 
To train the proposed models, we chose to minimize MAE instead of MSE since MAE is more interpretable in the case of $h$-index, and provides a generic and even measure of how well our model is performing.
Therefore, for very large differences between the $h$-index and the predicted $h$-index (\eg $y=120$ vs. $\hat{y}=40$), the function does not magnify the error.

With regards to the different types of features, we can see that the graph metrics alone correspond to a much stronger predictor of the future $h$-index compared to the features extracted from the papers' textual content.
Due to the rich information encoded into these features, MLP achieves similar performance to that of the two GNN models.
This is not surprising since GNN and GCN consist of two message passing layers followed by a multi-layer perceptron which is identical in terms of architecture to MLP.
The features extracted from the authors' papers seem not to capture the actual impact of the author.
It is indeed hard to determine the impact of an author based solely on the textual content of a subset of the papers the author has published.
Furthermore, these features have been produced from a limited number of an author's papers, and therefore, they might not be able to capture the author's relationship with similar authors because of the diversity of scientific abstracts and themes.
GCN is in general the best-performing method and yields the lowest levels of MAE.
Overall, given the MAE of $2.57$, we can argue that GNNs can be useful for semi-supervised prediction of the authors' future $h$-index in real-world scenarios.

\begin{table}[t]
    \centering
    \def\arraystretch{1.4}
    \begin{tabular}{l|c|c|c} \hline
    \multirow{2}{*}{Author} & \multirow{2}{*}{$h$-index} & \multicolumn{2}{|c}{Predicted $h$-index} \\ \cline{3-4}
    & & GCN & GNN \\ \hline
    Jiawei Han & 131 & 94.75 & 115.09 \\
    Jie Tang & 45 & 30.71 & 33.74 \\ 
    Yannis Manolopoulos & 40 & 43.79 & 49.63 \\ 
    Lada Adamic & 48 & 30.67 & 22.71 \\
    Zoubin Ghahramani & 77 & 40.64 & 43.33 \\
    Michalis Vazirgiannis & 36 & 25.76 & 27.89 \\
    Jure Leskovec & 68 & 37.77 & 38.26 \\
    Philip S Yu & 120 & 100.42 & 119.93 \\
    \hline
    \end{tabular}
    \caption{The actual $h$-index of a number of authors and their predicted $h$-index.}
    \label{tab:results_authors}
\end{table}

To qualitatively assess the effectiveness of the proposed models, we selected a number of well-known scholars in the field of data mining and present their $h$-index along with the predictions of our two models in Table~\ref{tab:results_authors}.
Keeping in mind that the overwhelming majority of authors have a relatively small $h$-index (as has been observed in Figure~\ref{fig:hindex}), it is clear that these are some of the most extreme and hard cases which can pose a significant challenge to the proposed models.
Even though the objective function of the proposed models is MAE which does not place more weight on large errors (which can happen for authors that have high $h$-index values), still, as can be seen in Table~\ref{tab:results_authors}, the models' predictions are relatively close to the actual $h$-index values of the authors in most cases.

\section{Conclusion}\label{sec:conclusion}
In this paper, we developed two GNNs to deal with the problem of predicting the $h$-index of authors based on information extracted from co-authorship networks. 
The proposed models outperform standard approaches in terms of mean absolute error in a semi-supervised setting, \ie when the training data is scant or not trustworthy, which is often the case in bibliographic data.
We also underline the flexibility of our model which combines the graph structure with the textual content of the authors' papers and metrics extracted from the co-authorship network.
In the future, we plan to investigate how multiple paper representations can be aggregated in the model, instead of computing the average of the top ones.
Moreover, we will evaluate the performance of our models in other fields such as in physics and in chemistry.

\input{references}
\end{document}

%% file: references.tex
%
%
\bibliographystyle{spmpsci.bst}
\bibliography{biblio}